# A Type of Virtual Force based Energy-hole Mitigation Strategy for Sensor Networks

Chao Sha, Chunhui Ren, Reza Malekian, *Senior member*, *IEEE*, Min Wu, Haiping Huang, *Member*, *IEEE*, Ning Ye

*Abstract*—In the era of Big Data and Mobile Internet, how to ensure the terminal devices (e.g., sensor nodes) work steadily for a long time is one of the key issues to improve the efficiency of the whole network. However, a lot of facts have shown that the unattended equipments are prone to failure due to energy exhaustion, physical damage and other reasons. This may result in the emergence of energy-hole, seriously affecting network performance and shortening its lifetime. To reduce data redundancy and avoid the generation of sensing blind areas, a type of Virtual Force based Energy-hole Mitigation strategy (VFEM) is proposed in this paper. Firstly, the virtual force (gravitation and repulsion) between nodes is introduced that makes nodes distribute as uniformly as possible. Secondly, in order to alleviate the "energy-hole problem", the network is divided into several annuluses with the same width. Then, another type of virtual force, named "virtual gravity generated by annulus", is proposed to further optimize the positions of nodes in each annulus. Finally, with the help of the "data forwarding area", the optimal paths for data uploading can be selected out, which effectively balances energy consumption of nodes. Experiment results show that, VFEM has a relatively good performance on postponing the generation time of energy-holes as well as prolonging the network lifetime compared with other typical energy-hole mitigation methods.

## I. INTRODUCTION

Nowadays, a large number of intelligent terminals have been widely deployed in our living environment. It seems that the viewpoint, "every grain of sand is a computer", proposed by Mark Weiser [1] is gradually becoming a reality. However, the energy issue is still an important fetter in the whole Internet of Things, especially in the development of Wireless Sensor Networks (WSNs). Since the battery powered nodes are constrained in energy resource, it is crucial to prolong the network lifetime of sensor network [2]. On the other hand, many important properties of the network such as coverage, connectivity, data fidelity, and lifetime are also affected by the way nodes are deployed [3-4]. In general, nodes are deployed randomly or in a pre-planned manner in an area called ROI (Region Of Interest). Whatever the deployment method is adopted, once the network starts to run, nodes may die due to hardware failures (including physical damage), energy depletion, or even a small embedded software bug, resulting in energy-holes [5]. The so-called "energy-hole" is an area that can no longer be perceived by any node (apparently, the nodes that could have monitored the area are dead). In theory, the generation time and position of energy-holes are often uncertain in a network with randomly deployed nodes. If there are no energy replenishment, the energy-holes will inevitably appear, and their scope will continue to expand.

On the other hand, as energy consumption is exponentially increased with the communication distance according to the energy consumption model, multi-hop communication is beneficial to data gathering for energy conservation. However, such a network structure is not without problems. Since the nodes close to the Sink need to forward data from other nodes in the same cluster, they exhaust their energy quickly, leading to an energy-hole around the Sink [6]. In this case, the network will be gradually divided into more and more areas that can not communicate with each other, resulting in rapid failure of the entire system. Wu *et al*. [7] also proposed that the lifetime of a uniformly deployed sensor network is mainly limited by the availability of nodes around the Sink.

No more data can be delivered to the Sink after an energy-hole appears [8-9]. In addition, nodes near these energy-holes are required to bear the data load of those death nodes so that the energy consumption level of them increases more rapidly [10-12]. More importantly, all these phenomena happen earlier than we expected. As a result, when the energy-holes appear, the residual energy of most of the alive nodes is relatively high. How to make full use of these nodes and do everything possible to postpone the generation time of energy-hole as well as to mitigate the adverse effects of it is worth studying. Perillo *et al*. [13] first analyzed in what condition the energy-holes appears. Olariu *et al*. [14] then proved that the energy-hole inevitably appear in sensor network. Many researchers regard that most of the energy-holes locate around the Sink [15-16]. So, they have put forward some energy-efficient routing protocols to try to balance work load between nodes. Moreover, energy consumption balance is also a key metrics impacting on the performance of sensor network [17]. One of the most efficient methods to achieve energy balance is to optimize the deployment and configuration of the network. In addition, the sleep scheduling and the alternate

Manuscript received February XX, 2019; revised XX XX, 20XX; accepted XX XX, 20XX. Date of publication XX XX, 20XX; date of current version XX XX, 20XX. This work was supported in part by the National Natural Science Foundation of P.R. China (61872194, 61872196), Jiangsu Natural Science Foundation for Excellent Young Scholar (BK20160089), Six Talent Peaks Project of Jiangsu Province (JNHB-095), "333" Project of Jiangsu Province, Qing Lan Project of Jiangsu Province, Innovation Project for Postgraduate of Jiangsu Province (SJCX18_0295) and 1311 Talents Project of Nanjing University of Posts and Telecommunications.
C. Sha, C. H. Ren, M. Wu, H. P. Huang and N. Ye are with the School of Computer Science, Software and Cyberspace Security, Nanjing University of Posts and Telecommunications, Nanjing, Jiangsu, China, 210003 (e-mail: shac@njupt.edu.cn)
R. Malekian is with the Department of Computer Science and Media Technology, Malmö University/Internet of Things and People Research Center, Malmö University, Malmö, 20506, Sweden working strategies for nodes are also effective ways to postpone the generation time of energy-holes and prolong the network lifetime. However, if the distribution of nodes is non-uniform, the implementation effect of the above strategies will be greatly reduced. Therefore, "energy-hole mitigation" is in fact a multi-stage collaborative optimization process.

## II. Related Works

With the advent of the era of super-large-scale ubiquitous interaction, how to maximize system lifetime and balance network energy consumption without human intervention has become the most urgent problem in WSN. In order to alleviate the energy-hole problem, scholars have carried out many targeted researches from the following aspects.

### A. Non-uniform Node Distribution on Mitigating the Energy-hole Problem

In WSNs, the energy consumption on transmission is nearly proportional to the fourth power of the hop distance, when it is large. For this reason, Jia *et al.* proposed an optimal deployment scheme for the relay nodes to solve the problem of excessive energy consumption on long-distance data uploading in large-scale networks [18]. Near the center of the network, the node whose distance from the Sink meets certain criteria was regarded as a relay. When the residual energy of the edge node was lower than the threshold, it chose the most suitable relay node to forward its data. Obviously, the selection strategy of relay nodes directly effects the generation time of energy-holes.

In [19], Nguyen *et al.* accurately calculated out the position and area of the energy-hole. Then, the outer polygon of this energy-hole was constructed, and the active nodes around this polygon all received the message about this energy-hole. As a result, data packets can be transmitted along the periphery of this energy-hole, which minimizes the packet loss rate and maintains the throughput of the network. However, this method increases the data transmission delay.

A type of three-layer based network consisting of a static Sink, some static sensor nodes and several proxy nodes was proposed in [20]. In order to achieve energy balance, the initial energy of these two types of nodes was set to different values. With the help of the mobile proxy nodes deployed randomly near the network center, this method effectively postpones the generation time of energy-holes near the static Sink. However, in areas far from the center, there were still many dead nodes. With the deepening of research, people gradually find that the relationship between energy-holes and network coverage scheduling is more obvious. For example, for a static network, its coverage often depends on how to reasonably select several optimal and minimum active node sets. Only the nodes in one of the active node sets need to monitor the network while the rest of them can sleep, which greatly reduces energy consumption of the whole network. Sun *et al.* pointed out that as long as the node sets are reasonably selected, it can ensure that the energy-holes do not appear for a long time [21].

Furthermore, Kacimi *et al.* discussed the load balancing techniques to mitigate energy-hole problem in large-scale WSNs, and proposed a distributed heuristic solution to balance energy consumption of nodes by adjusting their transmission power [22]. However, it is unrealistic to frequently change the data transmission power of nodes. Moreover, this is also easy to increase the energy consumption of nodes. While Liu *et al.* pointed out that the deployment density of nodes in the network should be inversely proportional to the distance between them and Sink [23]. On the basis of this theory, they accurately calculated out the First Node Die Time (FNDT) and All Nodes Die Time (ANDT). This is also the basis of the network model proposed by our algorithm.

### B. Achieving Energy Consumption Balance with the help of Mobile Sink

Recently, many studies have shown that the energy-hole problem can be effectively mitigated by using one or more mobile Sinks [24].

Zhang *et al.* [25] proposed an optimal cluster-based strategy to achieve load balancing in data collection with the help of several mobile Sinks. Moreover, the Rendezvous Points (RPs) and Rendezvous Nodes (RNs) were then introduced to further reduce time delay. However, the network topology in this method can never be changed anymore, which is often unrealistic.

To enhance the network coverage and minimize the sensing redundancy, Sahoo *et al.* proposed a distributed energy-hole repair method [26]. According to their locations, nodes were classed into three categories, that were cross triangle node, hidden cross triangle node, and non-cross triangle node. Through the interaction between different types of nodes, high coverage redundancy areas and energy-holes were found out. Subsequently, several nodes located in the high coverage redundant area were moved to the location of the energy-holes to repair them. Obviously, this method can reduce the probability of energy-holes to some extent by adjusting the network topology, but its real-time performance still needs to be improved.

In [27], the author tried to solve the energy-hole problem by using some mobile relay nodes. The network is divided into several clusters with different sizes, and each "relay node" is responsible for uploading the sensing data of several clusters to Sink. That is to say, the relay node is a mobile data collector relative to the "sub-region" composed of several clusters. This algorithm is very suitable for event-driven networks or networks that need continuous data transmission. With the help of these "mobile relay nodes", the energy efficiency as well as load balance of the whole network have been greatly improved. However, this method relies too much on "relay nodes". Once these relay nodes fail or their data forwarding efficiency is low, some nodes (especially the cluster header nodes) will prematurely die, resulting in the energy-holes.

Abo-zahhad *et al.* used the clustering mechanism combined with one mobile Sink to alleviate the energy-hole problem [28]. Based on the adaptive immune algorithm of energy dissipation, they calculated the optimal number of cluster heads. Then, the trajectory of the mobile Sink was also obtained, which can reduce the burden of cluster heads to a certain extent and postpone the generation time of energy-holes. However, this method did not fundamentally solved the problem of unbalanced energy consumption of nodes.

In the era of mobile Internet, more and more scholars point out that besides Sink or relay nodes, other nodes should also be able to move in WSNs. They believe that after being deployed, nodes can move freely to the area where the energy-holes appear, so that those holes can be repaired. In this case, how to select out the appropriate nodes to repair the holes and how to move these nodes to the right position are two critical issues. One of the feasible solutions is to use the "virtual force". Adjacent nodes can exert attractive or repulsive force to each other. For non-adjacent nodes, there is no virtual force between them. However, most of the current researches on virtual force are still in the conceptual stage. In these works, the magnitudes of virtual force between nodes are not quantified according to the specific network structure. In this paper, the distance between nodes in specific network structure is fully analyzed, and two kinds of quantification models for the virtual force are described. On this basis, the network topology is optimized that can mitigate the energy-hole problem.

*C. Mitigating the Energy-hole Problem in Circular Network*

Although the shape of Wireless Sensor Network is various in real scenarios, it can be abstracted as a circular network whether it is a cluster based structure or a multi-hop network structure. In this section, we introduce the energy-hole mitigation strategies in the circular network.

In [29], a circular network is divided into several annuluses at first, and then each annulus is divided into smaller sectors. The authors proved that this logical partition method can reduce the probability of energy-holes. However, this method did not take into account the distance between the node and cluster head in the real scene, which may result in high energy consumption.

A type of Wireless Sensor Network Energy Hole Alleviating (WSNEHA) algorithm was proposed in [30]. The network was divided into several annuluses with the same width, as shown in Figure 1. A static Sink was located in the center of the network, and nodes were evenly distributed in each annulus. For each node in the second annulus, according to the real-time energy consumption rate of all its possible successors, the most suitable one was selected as its forwarding node. Thus, the load of each node in the innermost annulus can be balanced. On this basis, Jan *et al*. [31] further discussed the load balancing problem of nodes in each annulus, and its network structure is shown in Figure 2. In this model, nodes were randomly distributed in each annulus. Based on distance and energy constraints, each node adopted the minimum bit error rate transmission strategy to select its next hop forwarder. However, neither method mentioned above is flexible enough. In the case of dense deployment of nodes, their effect on mitigating energy-holes is not obvious.

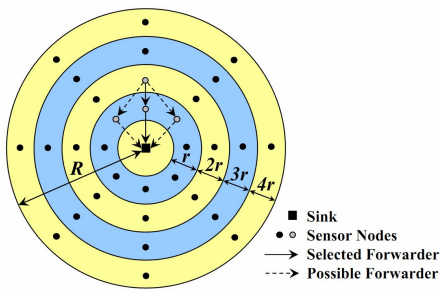

Fig. 1. Network model of WSNEHA [30].

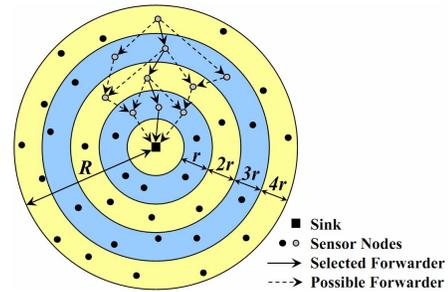

Fig. 2. Network model proposed in [31].

Wu *et al*. [7] pointed out that in a circular network with non-uniform distribution of nodes and constant data uploading rate, the phenomenon of unbalanced energy consumption of nodes is unavoidable. However, if the number of nodes can be increased from the outer annulus to the inner one layer by layer according to geometric series, the energy-hole problem may be avoided to some extent. For this reason, they proposed a non-uniform node deployment strategy and obtained the proportional relationship between the number of nodes in the adjacent annulus. However, there were too many nodes deployed near the network center, which easily causes a lot of coverage redundancy.

*D. Using Energy Replenishment Technology to Prevent the Appearance of Energy-holes*

All the methods mentioned above can not fundamentally eliminate the energy-hole, and they can only postpone the time it appears. In recent years, the continuous improvement of Wireless Energy Transfer (WET) technology has made it possible to replenish energy to the ubiquitous sensor nodes by a non-contact way, that is, Wireless Rechargeable Sensor Networks (WRSNs). For most of the energy replenishment strategies, one or more Mobile Wireless Charger (MWC) are often employed to periodically recharge all nodes along some fixed trajectories, or they only recharge the nodes on demand when randomly walking. In this way, as long as the nodes are able to be charged in time before death, the energy-hole can be eliminated theoretically.

Wang *et al*. [32] divided the network into annuluses to alleviate the "energy-hole problem". The recharging threshold was set for each node depending on its energy consumption rate and the length of the request queue, which is more reasonable for large-scale networks. Moreover, the recharging order of nodes was obtained by constructing a Minimum cost Spanning Tree (MST) covering all nodes in the recharging request queue. In this way, they can achieve a relatively high recharging profit. However, the recharging threshold in all these preceding works was mainly determined by node's energy, which failed to intuitively reflect the residual lifetime of a node.

Zhu *et al*. [33] strictly limited the battery capacity of WCV and proposed a type of Node Failure Avoidance Online Charging scheme (NFAOC) based on node's real-time energy consumption rate. Nodes which cause the smallest number of dead nodes is selected as the next recharging object, thereby assuring high recharging efficiency. It should be pointed out that this strategy tends to fall into local optimum during the path construction phase, which reduces the number of nodes it can serve. Therefore, it can not completely eliminate the energy-hole.

In order to make MWC charge more nodes in a limited time, Xu et al. [2] assumed that the actual energy required by the node should be evenly replenished during multiple recharging rounds. That is to say, only a part of energy need to be replenished to node each time. Although this method increases the number of nodes that a MWC can serve, it also accelerates the scheduling frequency as well the energy consumption rate of the MWC.

Although WRSN plays a positive role in alleviating the energy-hole problem, it is also undeniable that the cost of the wireless mobile charger is relatively higher and the scheduling algorithm is a little complex. In addition, the moving speed as well as the charging efficiency of WMC are low. All these factors make it impossible for such methods to completely prevent the generation of energy-holes.

On the basis of the above researches, we propose an energy-hole mitigation strategy with the help of two kinds of virtual force. Contributions of this paper can be concluded as follows.

Firstly, under the action of virtual gravitation and repulsion force, nodes are uniformly distributed in the network which ensures that no blind area of perception appear during a long time.

Secondly, position of each node is further optimized with the help of the "virtual gravity generated by annulus". This not only reduces the load of nodes near the center but also ensures the coverage of the whole network.

Finally, nodes are classified into two categories, which are responsible for sensing and data forwarding, respectively. The optimal number of each kind of nodes located in each annulus and the weights of all possible data uploading paths in the "data forwarding area" are calculated out for path selection. This further balance energy consumption on data uploading.

The remainder of this paper is organized as follows. The related works are described in section II. And the network model as well as the virtual force based energy-hole mitigation method are described in section III. Experimental results of VFEM are shown in section IV and the conclusion is provided in the last section.

## III. METHOD DESCRIPTION

### A. Network Model

It is assumed that $N$ sensor nodes are randomly deployed in a circular region whose radius is $R$. The base station $B$ is located at the center of network. According to [7], both the maximum communication radius and maximum sensing radius of each node are the same (the value of them is defined as $r$). The initial topology after deployment is shown in Figure 3, in which the sensing data can be sent to the base station via one-hop or multi-hop transmission.

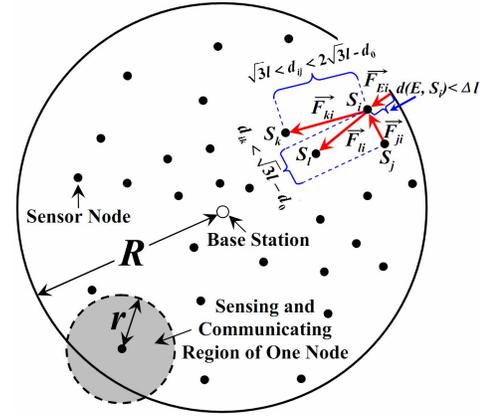

Fig. 3. Nodes are randomly deployed in the circular network.

Without loss of generality, the energy dissipation model adopted by VFEM is the same as that in [7] and [34], as shown in Figure 4.

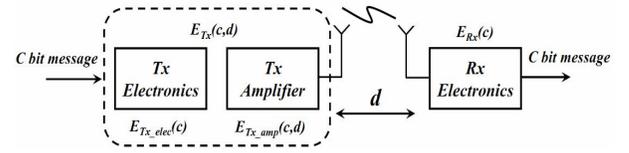

Fig. 4. Energy dissipation model [30].

In formula (1) and (2), $E_{send}$ and $E_{rec}$ are the energy consumption of one node during its sending and receiving phase. $E_{elec}$ is the unit energy consumption of the circuit. $\varepsilon_{fs}$ and $\varepsilon_{amp}$ are the signal amplifier in the free space and multi-path fading environment, while $d' = \sqrt{\varepsilon_{fs}/\varepsilon_{amp}}$ denotes the threshold distance.

To transmit a $c$-bit message a distance $d$, the radio expends

$$E_{send}(c,d) = \begin{cases} cE_{elec} + c\varepsilon_{fs}d^2 & d < d' \\ cE_{elec} + c\varepsilon_{amp}d^4 & d \geq d' \end{cases} \quad (1)$$

And to receive this message, the radio expends

$$E_{rec}(c) = cE_{elec} \quad (2)$$

### B. Node Position Adjustment Based on Virtual Force between Sensors

Generally speaking, random deployment is easy to realize in sensor network. However, uneven distribution of nodes in the network could easily cause data redundancy and perceptual holes. On the other hand, the long hop distance between nodes may cause unbalanced energy consumption on communication, and this also shorten the network lifetime. Thus, the primary goal of VFEM is to make nodes distribute as uniformly as possible.

According to [28], if sensor nodes can be finally deployed as the structure shown in figure 5, the network is regarded as to achieve a nearly uniform distribution. The black dots are the ideal positions of nodes, while the hexagon in this figure is the actual sensing region of each node when no coverage hole appears. We can see that the area composed by all the regular hexagons cover the whole network. In this case, the density of nodes (defined as $\rho$) is $2/3\sqrt{3}l^2$ [28]. $l$ is the length of each regular hexagon and it is not difficult to know that, $l = \sqrt{2\pi/3\sqrt{3}(N+1)}R$.

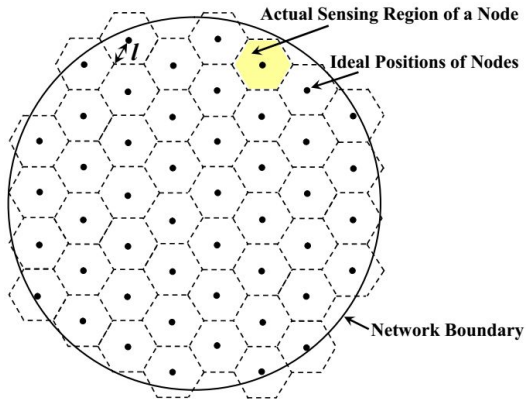

Fig. 5. Nodes are uniformly deployed in the circular network.

Nowadays, people have made a lot of efforts on improving both the distribution of nodes in network and optimizing the network topology [7, 13]. Node's position adjustment with the help of virtual force is one of the effective strategies. The concept of virtual force was first proposed in path planning and barrier avoidance for robots. Then, it was introduced to Wireless Sensor Networks to try to solve the coverage enhancement problem. The basic idea of virtual force is to assume each node as charge so that it will obtain virtual force from other nodes. With the effect of this virtual force, nodes may move to other positions and finally achieve equilibrium on force which ensures full coverage of the network.

Thus, in VFEM, it is regarded that one node has virtual force (gravitation and repulsion) with another one when the distance between them meets certain constraints. With the help of these virtual force, nodes can move within the network (there is friction between node and ground) and finally reach to a stationary state in which the distance between two adjacent nodes is nearly equal to $l$. Here, we take two sensor nodes ($S_i$ and $S_j$) as an example to describe the definition of virtual force in VFEM.

1) When $\sqrt{3}l - d_0 \leq d_{ij} \leq \sqrt{3}l$, it is assumed that, there is no force between $S_i$ and $S_j$. $d_{ij}$ is the Euclidean distance between $S_i$ and $S_j$. Moreover, to avoid nodes' repeated movement that is caused by virtual force, $d_0$ is introduced as the buffering distance and $d_0 \in (0, \sqrt{3}l/2)$.

2) If $d_{ij} < \sqrt{3}l - d_0$, it is obvious that, the distance between $S_i$ and $S_j$ is short. To make the nodes approach the state of uniform distribution as shown in Figure 5, $S_i$ and $S_j$ need to get away from each other. Thus, there is only repulsion force between $S_i$ and $S_j$ in this case, and the value of this force is $\eta/d_{ij}^\beta$. $\eta$ is defined as the coefficient of this repulsion and $\beta$ is an adjustable parameter. Values of $\eta$ and $\beta$ are described in the experimental section.

3) When $\sqrt{3}l < d_{ij} < 2\sqrt{3}l - d_0$, the distance between $S_i$ and $S_j$ could be considered a little long. Similarly, to make the nodes approach the state of uniform distribution, $S_i$ and $S_j$ need to get close to each other. Therefore, there is only gravitation force between $S_i$ and $S_j$ in this case, and the value of this it is set to $\lambda d_{ij}^\beta$. $\lambda$ is the coefficient of this gravitation.

4) If $d_{ij} \geq 2\sqrt{3}l$, it is known that, the distance between $S_i$ and $S_j$ is too long. So, there is no force between these two nodes.

We use $F_{ji}$ to mark the virtual force from $S_j$ to $S_i$. In summary, we can get

$$F_{ji} = \begin{cases} \eta/d_{ij}^\beta & d_{ij} < \sqrt{3}l - d_0 \\ \lambda d_{ij}^\beta & \sqrt{3}l < d_{ij} < 2\sqrt{3}l - d_0 \\ 0 & \sqrt{3}l - d_0 < d_{ij} \leq \sqrt{3}l \ \ or \ \ d_{ij} \geq 2\sqrt{3}l \end{cases} \quad (3)$$

It is worth mentioning that, the base station is regarded as a common sensor node, and it generates repulsion and gravitation to other nodes. However, to ensure that the location of $B$ does not get change, we assume that the base station is not acted upon by the virtual force generated by other nodes.

What's more, to avoid nodes moving out of the network, the network boundary exists repulsive force to $S_i$ (marked as $F_{bi}$) in a certain range. The direction of $F_{bi}$ points to the center of network and the value of it is defined as follows.

$$F_{bi} = \begin{cases} \eta/(d(b,S_i))^\tau & 0 \leq d(b,S_i) \leq \Delta l \\ 0 & d(b,S_i) > \Delta l \end{cases} \quad (4)$$

In formula (4), $d(b,S_i)$ is the shortest distance between $S_i$ and the network boundary. Thus, $d(b,S_i)=R-d(B,S_i)$. ($d(B,S_i)$ is the Euclidean distance between base station $B$ and node $S_i$). In addition, $\Delta l$ is an adjustable distance, and $\Delta l << l$.

Moreover, from the above analysis, it is easy to know that, the minimum gravitation and repulsion force between two nodes are $\lambda(\sqrt{3}l)^\beta$ and $\eta/(\sqrt{3}l - d_0)^\beta$, respectively. The minimum repulsion force generated from the network boundary to the node is $\eta/(\Delta l)^\tau$. Therefore, to make the nodes approach the state of uniform distribution, the friction $f$ ought to be smaller than the minimum value of virtual force. That is,

$$f < Min\left(\lambda(\sqrt{3}l)^\beta, \eta/(\sqrt{3}l - d_0)^\beta, \eta/(\Delta l)^\tau\right) \quad (5)$$

Thus, the resultant force acting on $S_i$ could be expressed as follows.

$$\overrightarrow{F(S_i)} = \sum_{j=1, j \neq i}^{N} \overrightarrow{F_{ji}} + \overrightarrow{F_{bi}} + \overrightarrow{f} \quad (6)$$

So, when the sensor node is deployed in the network, it moves along the direction of the resultant force until the value of $\overrightarrow{F(S_i)}$ is equal to zero and will not change any more.

### C. Optimum Number of Nodes in Each Annulus

Although the uniform distributed nodes can ensure that no perceptual blind area appears for a long time, when radius of the network is too big, nodes have no choice but to transmit data to the base station with multi-hop transmission manner. In this case, it will increase the load of the nodes near the center. To relieve the "energy-hole problem", the "virtual gravity generated by annulus" is proposed to further optimize the positions of nodes.

Similar to Wu [7] and our previous work [35], in VFEM, the circular network is divided into $k$ annular regions with the same width (marked as $d_w$), as shown in Figure 6. The annular regions from inside to outside are marked as $C_0$, $C_1$…$C_{k-1}$. Actually, $C_0$ is a circle whose radius is $d_w$. As mentioned above, nodes near the center may bear more data transmission tasks. So, it is a good choice to allocate different number of nodes in different annular regions to balance energy consumption. However, this may increase the coverage redundancy and the efficiency on data collection may also be reduced. On the other hand, in most data collection and routing methods, nodes are often regarded to be "fully functional", that is, they can act as sensing terminals, relay nodes, or even cluster heads. Nevertheless, the difference on energy consumption between nodes may become larger and larger in this way. Therefore, it is assumed that there are two kinds of nodes exist in the network.

*Sensing Nodes*: These nodes (the white dots in Figure 6) only do the sensing and data uploading tasks, and they can not receive data form other nodes.

*Relay Nodes*: This type of nodes (the grey dots in Figure 6) only receive data sending from nodes in the adjacent outer annulus. Then, they forward these data to the relay nodes located at the adjacent inner annulus so that all the sensing data can finally be transmitted to the base station. It is worth noting that, relay nodes can not perceive information.

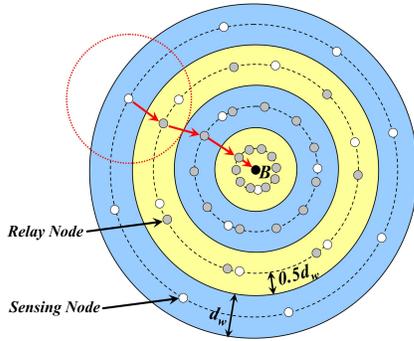

Fig. 6. Network model of VFEM.

Moreover, to further enhance the efficiency on sensing and to facilitate the relay nodes finding out the optimal data uploading paths, nodes in the same annulus should be located at the center of this annulus (the dotted line in Figure 6). In this case, the shortest distance from the node to both the outer and inner side of the annulus is $d_w/2$.

It is assumed that, in a round of data collection time, each sensing node gets $c$ bits data and forwards them to the next-hop node until to the base station. To avoid the "energy-hole problem", energy consumption rate of each node should be roughly the same with each other. That is, $E_0'/E_0 \approx E_1'/E_1 \approx \cdots E_m'/E_m \cdots \approx E_{k-1}'/E_{k-1}$. $E_m'$ is defined as the total energy consumption of nodes in $C_m$ during a round of data gathering time, while $E_m$ is the sum of initial energy of all the nodes in $C_m$. It is also assumed that, $N_m^s$ and $N_m^r$ represent the number of sensing nodes and relay nodes in $C_m$, respectively. Thus,

$$\frac{e_t N_{k-1}^s}{e_0 N_{k-1}^s} \approx \frac{e_t N_{k-2}^s + (e_t + e_r)\sum_{i=k-1}^{k-1} N_i^s}{(N_{k-2}^s + N_{k-2}^r)e_0} \approx \cdots$$
$$\approx \frac{e_t N_1^s + (e_t + e_r)\sum_{i=2}^{k-1} N_i^s}{(N_1^s + N_1^r)e_0} \quad (7)$$
$$\approx \frac{e_t' N_0^s + (e_t' + e_r)\sum_{i=1}^{k-1} N_i^s}{(N_0^s + N_0^r)e_0}$$

In formula (7), $e_0$ represents the initial energy of each node. $e_t$ and $e_t'$ are the transmission energy consumption of nodes in $C_m(m>0)$ and $C_0$ respectively during a round of data collection time. $e_r$ is the energy consumption on data receiving. That is to say,

$$e_t = cE_{elec} + c\varepsilon_{fs}d^2 \quad (8)$$
$$e_t' = cE_{elec} + c\varepsilon_{fs}(d_w/2)^2 \quad (9)$$
$$e_r = cE_{elec} \quad (10)$$

It is necessary to point out that, the parameter $d$ in formula (8) is defined as the expected value of distance between a pair of transmitting and receiving nodes located in adjacent annuluses. The value of it is discussed in the following section. As mentioned before, nodes in $C_0$ only need to send data to the base station, so the distance about the last hop is $d_w/2$.

From the above analysis, it is known that, the number of the sensing and relay nodes in each annulus are critical for balance of energy consumption. Therefore, the values of $N_m^s$ and $N_m^r$ are discussed as follows.

To enhance the efficiency on data collection, sensing nodes should meet the following conditions.

1) The sum of sensing regions of all these nodes should cover the whole network. That is to say, there is no blind area on sensing.

2) Each sensing node should be able to upload data to at least one relay node in the adjacent inner annulus.

3) The proportion of overlapping coverage area in the network should be as small as possible.

In order to meet condition 1) and 2), both the sensing radius and communication radius are set to $1.5d_w$. Meanwhile, to reduce the redundancy on data collection, nodes in each annulus should be uniformly distributed. Moreover, area of the overlapping sensing region should be as small as possible, as shown in Figure 7. The red dashed lines in Figure 7 are the perception boundaries of each sensing node in $C_m$.

According to cosine theorem, the relationship between the three sides of △ABC in Figure 7 can be expressed as follows.

$$(1.5d_w)^2 = ((m+0.5)d_w)^2 + ((m+1)d_w)^2 \quad (11)$$
$$- 2(m+0.5)(m+1)d_w^2\cos\alpha$$

Thus,

$$N_m^s = \left\lceil \frac{\pi}{\arccos\left((2m^2+3m-1)/(2m^2+3m+1)\right)} \right\rceil \quad (12)$$

And the value of $N_m^r$ could also be solved out by formula (7) and (12).

$$N_m^r = \begin{cases} \left\lceil (1+e_r/e_t)\sum_{i=m+1}^{k-1} N_i^s \right\rceil & m \in [1, k-2] \\ \left\lceil (1+e_r/e_t')\sum_{i=1}^{k-1} N_i^s \right\rceil & m = 0 \end{cases} \quad (13)$$

That is,

$$N_m^r = \begin{cases} \left\lceil \left(1+\frac{E_{elec}}{E_{elec}+\varepsilon_{fs}d^2}\right)\sum_{i=m+1}^{k-1} N_i^s \right\rceil & m \in [1, k-2] \\ \left\lceil \left(1+\frac{4E_{elec}}{4E_{elec}+\varepsilon_{fs}d_w^2}\right)\sum_{i=1}^{k-1} N_i^s \right\rceil & m = 0 \end{cases} \quad (14)$$

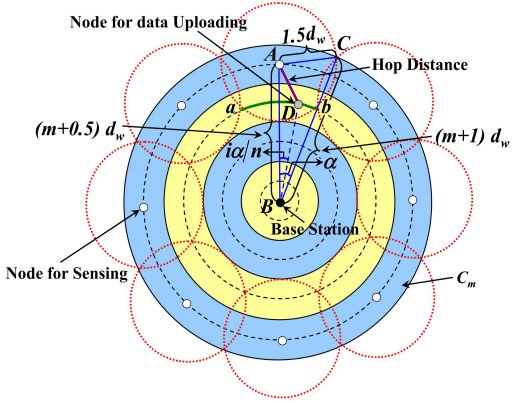

Fig. 7. Deployment of nodes with no energy-holes.

On the other hand, for the sensing node in $C_m(m>0)$, the shortest and longest distance from this node to its next-hop successor are $d_w$ and $1.5d_w$ respectively. As shown in Figure 7, the next-hop successor of node $A$ (e.g., node $D$) can only be located at one position on curve $ab$. The angle value of $\angle ABD$ is set to $i\alpha/n$ ($i \leq n$). Thus, by using the cosine theorem, it is easy to know the distance of segment $AD$, and the length of the single hop distance could also be calculated out by formula (15).

$$d = \lim_{n \to \infty} \left( \frac{1}{n} \sum_{i=1}^{n} \sqrt{\begin{array}{c}((m+0.5)d_w)^2 + ((m-0.5)d_w)^2 \\ -2(m+0.5)(m-0.5)d_w^2 \cos(i\alpha/n)\end{array}} \right) \quad (15)$$

It is known from formula (12) that, in formula (15),
$$\alpha = \arccos\left((2m^2+3m-1)/(2m^2+3m+1)\right) \quad (16)$$

### D. Nodes' Positions Optimization based on Virtual Force Generated from Annulus

As mentioned earlier, only if the number of the sensing and relay nodes in $C_m$ are equal to $N_m^s$ and $N_m^r$ respectively, energy consumption of them are nearly the same with each other. Thus, we assume that the curve located in the middle of each annulus (e.g., $C_m$) can also generate virtual force (marked as $\overline{F(C_m)}$) so that corresponding number of nodes will finally move onto each curve. The final distribution of nodes is shown in Figure 6.

Sphere of influence from this virtual force generated by curve in $C_m(m>0)$ is also an annular region whose center is the base station. The inner and outer boundaries of these regions are drawn with the red solid lines in Figure 8, and the width of each region is marked as $d_F(C_m)$. So, lengths of the inner and outer diameters of $C_m$ are $2\sum_{i=0}^{m-1} d_F(C_i)$ and $2\sum_{i=0}^{m} d_F(C_i)$, respectively. The influenced area of the virtual force generated by the curve located in the middle of $C_0$ is a circle whose radius is $d_F(C_0)$.

As mentioned in Section III.B, all nodes have been nearly uniformly distributed in the network with the action of the virtual force between nodes. Now, the virtual force generated by annulus only needs to keep the number of nodes in the sphere of influence approximately equals to $N_m^s + N_m^t$. That is,

$$\pi \left( \sum_{i=0}^{m} d_F(C_m) \right)^2 (N/\pi R^2) = \sum_{j=0}^{m} (N_j^s + N_j^t) \quad (17)$$

So,

$$\sum_{i=0}^{m} d_F(C_m) = \sqrt{(R^2/N) \sum_{j=0}^{m} (N_j^s + N_j^t)} \quad (18)$$

The value of $d_F(C_m)$ can be calculated out by mathematical induction.

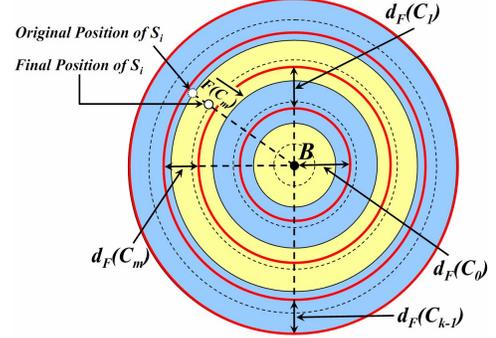

Fig. 8. Virtual force generated from annulus.

Due to this kind of virtual force, nodes move onto the curve located in the middle of each annulus. Take $S_i$ in Figure 8 as an example. The white dot formed by the dotted line is the position of $S_i$ after the action of the first kind of virtual force. We can see that this position is in the scope of influence of the virtual force generated by the dotted curve in $C_m$. Hence, $S_i$ further moves to the position of the white dot formed by the solid line.

Moreover, if nodes are uniformly distributed on the circular arc, the minimum redundancy can be guaranteed. Therefore, position of sensing nodes should be further improved. That is, for two adjacent sensing nodes (e.g., $S_i$ and $S_j$) in $C_m$, the value of $\angle S_i B S_j$ should be $2\pi/N_m^s$, as the white nodes shown in Figure 6.

In addition, from formula (13), it is not difficult to know that, the number of the sensing nodes located in $C_{m+1}$ ($m>0$) and $C_1$ satisfy the following constraints.

$$\Delta T_{avg} = \sum_{j=1}^{Num(R)} TP_j(\Delta t_{max}) / Num(R) \quad (19)$$

$$N_1^s < \left\lceil (1+e_r/e_t') \sum_{i=1}^{k-1} N_i^s \right\rceil = N_0^r \quad (20)$$

So, the number of the relay nodes in $C_m$ ($k-1>m\geq 0$) must be larger than that of sensing nodes in $C_{m+1}$. Thus, these relay nodes should also be uniformly distributed on the curve, as the gray nodes shown in Figure 6. This ensures that each sensing node is able to find out at least one relay node as its next-hop successor, as shown in Figure 7. Sensing nodes in $C_0$ straightly send data to the base station without forwarding.

Similarly, it is also known from formula (13) that, the number of relay nodes in $C_m$ ($k-2>m\geq 0$) is larger than that of relay nodes in $C_{m+1}$. That means when the relay nodes are uniformly distributed on each curve, each relay node in $C_{m+1}$ can certainly find out at least one relay node in $C_m$ as its next-hop successor.

Furthermore, with the help of formula (12) and (13), it can also be concluded that, when $m$ is smaller, the value of $N_m^s$ will be smaller too while the value of $N_m^r$ will be larger. That is, the closer to the base station, the less the number of sensing nodes, but the more the number of relay nodes. In this way, each node has more choices to select the optimal next-hop successor, and the "hotspot problem" can also be alleviated to a certain extent. It is helpful to postpone the generation time of energy-holes.

### E. Optimal Path for Data Uploading

Now, each node has moved onto the middle position of each annulus with the help of the virtual force generated from annulus, as shown in Figure 6. Meanwhile, all the sensing and relay nodes have been uniformly distributed on the curve. Although it is proved that, each node can find out at least one successor for data uploading, if there is no limitation on path selection, the hop distance might be long. As shown in Figure 9, the length of each hop is close to or equal to $1.5d_w$. This not only increases the energy consumption on communication, but also aggravates the burden on some relay nodes (e.g., node $S_i$).

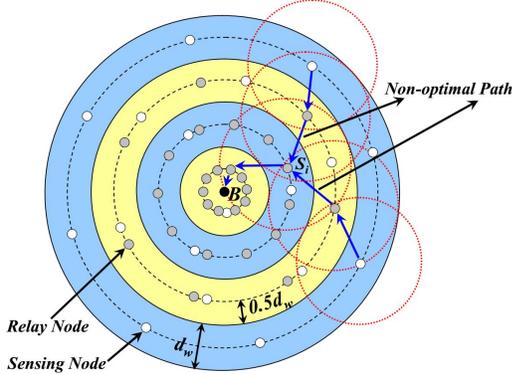

Fig. 9. Unreasonable data uploading paths.

Therefore, the concept of "data forwarding area" is proposed here. As we know, the boundary of sensing range of node $S_i$ in $C_m$ ($k>m>0$) will intersect with the curve in the middle of $C_{m-1}$ at two points, such as point $a$ and $b$ in Figure 7. The sector area formed by arc $ab$, line segment $aB$ and line segment $bB$ is defined as the "data forwarding area" (the sector region in Figure 10). For any sensing node $S_j$, the next-hop successor of it can only be selected out from the "data forwarding area".

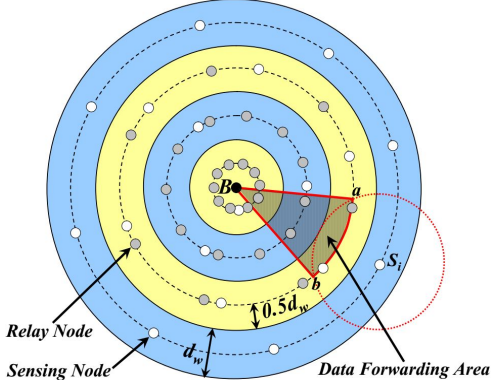

Fig. 10. Data forwarding area.

For one sensing node (e.g., $S_i$) in $C_{m+1}$ ($k-1>m>0$), it is assumed that $S_j$ (located at $C_j$) and $S_0$ (located at $C_0$) are two candidate relay nodes of $S_i$ in the "data forwarding area". Thus, "$S_i \rightarrow S_j \rightarrow S_{j-1} \rightarrow ... \rightarrow S_0 \rightarrow B$" can be regarded as one of the possible data uploading paths (marked as $path_i$). The weight of $path_i$ is defined as follows.

$$W(path_i) = \sum_{j=m}^{1} E_r(S_j)/d^2(S_j,S_{j-1}) + E_r(S_0)/d^2(S_0,B) \quad (21)$$

In formula (21), $E_r(S_j)$ is the residual energy of $S_j$. $d(S_j,S_{j-1})$ is the Euclidean distance between $S_j$ and $S_{j-1}$ while $d(S_0,B)$ is the Euclidean distance between $S_0$ and the base station. Sensing node $S_i$ calculates out all the possible paths' weights with the help of formula (21), and it then selects out the path with the largest value of weight as the data uploading path (the blue arrow in Figure 11).

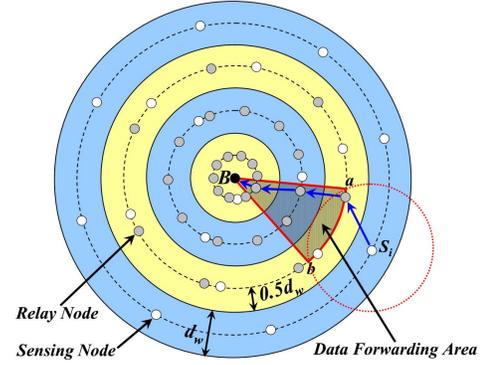

Fig. 11. The optimal data uploading path.

Both the residual energy of the relay nodes in "data forwarding area" and the hop distance between nodes are considered in selecting out the optimal data uploading path. For this reason, VFEM can ensure energy consumption balance to a certain extent. However, it is possible for some sensing nodes to choose the same relay node as the successor in their uploading paths, which greatly increases the energy consumption on some relay nodes (e.g., node $S_2$ in Figure 12). In order to solve this problem, each sensing node should monitor the residual energy of all the forwarding nodes in its "data forwarding area" at the end of each round of data uploading. When the residual energy of $S_j$ is lower than its threshold $E_r$ or the standard deviation for all the relay nodes in the "data forwarding area" is higher than the threshold $\delta$ (definition of $\delta$ is shown in formula (22) ), $S_j$ will select out another data uploading path according to formula (21). Meanwhile, these relay nodes whose residual energy is lower than $E_r$ are regard as dead nodes.

$$D(E_r(S_k)) = \frac{\sum_{k=1}^{Num(S_i)} E_r(S_k)^2}{Num(S_i)} - \left(\frac{\sum_{k=1}^{Num(S_i)} E_r(S_k)}{Num(S_i)}\right)^2 \quad (22)$$

In formula (22), $Num(S_i)$ is the total number of relay nodes in the "data forwarding area".

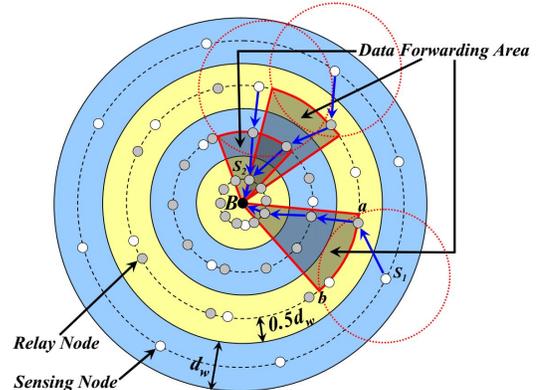

Fig. 12. Loads on the relay nodes.

It is worth mentioning that, in VFEM, the computation complexity on path selection is not high. Take a circular network which is divided into 5 virtual annuluses as an example, it is not difficult to know that, the number of sensing nodes in $C_4$, $C_3$, $C_2$, $C_1$ and $C_0$ are 11, 9, 7, 4 and 1. While the number of relay nodes in $C_3$, $C_2$, $C_1$ and $C_0$ are $\lceil 11(1+e_r/e_t) \rceil$, $\lceil 20(1+e_r/e_t) \rceil$, $\lceil 27(1+e_r/e_t) \rceil$ and $\lceil 31(1+e_r/e_t) \rceil$, respectively. Due to the fact that the sensing and relay nodes are uniformly distributed in each curve, there won't be too many relay nodes in each "data forwarding area". For example, for any sensing node in $C_4$, the total number of relay nodes in its "data forwarding area" is only $\lceil 11(1+e_r/e_t)/11 \rceil + \lceil 20(1+e_r/e_t)/11 \rceil + \lceil 27(1+e_r/e_t)/11 \rceil + \lceil 31(1+e_r/e_t)/11 \rceil$. Thus, the there are at most $\lceil (20 \times 27 \times 31/11^3)(1+e_r/e_t)^4 \rceil$ possible paths for selection. It is known from [30] that, $e_r$ is far less than $e_t$, so the number of the possible paths for data uploading is only about 13. Thus, computation cost on the optimal path selection algorithm in VFEM is low.

## IV. SIMULATION RESULTS AND ANALYSIS

Simulation results are carried out with the help of Matlab 8.5. All the experiments were carried on a server (the operating system is Win 10) with Intel Xeon (E3-1225V6) 3.3GHz CPU, 16GB memory(DDR4, 2400MHZ), 8MB cache and 2TB hard disk. All the algorithms were implemented via Java code. We compare VFEM with SNAA [35] (an energy-hole mitigation strategy proposed by our previous work) and the method proposed by Wu [7]. In SNAA, the circular network is also divided into virtual annuluses with the same width. Nodes are non-uniformly deployed, and the number of them increases in geometric progression from the outer annuluses to the inner ones, which effectively reduces the work load on nodes near the center. Moreover, each node can find its optimal parent by considering the residual energy of each candidate as well as the distance between the two nodes in adjacent annuluses. The node deployment model and the data uploading process in SNAA are shown in Figure 13 and 14. Values of the experimental parameters are shown in Table I. All the data of each experiment were obtained after 100 times of simulations.

TABLE I
PARAMETER VALUES

| Parameter | Symbol | Value | Unit |
|---|---|---|---|
| Network Radius | $R$ | 100 | $m$ |
| Number of Nodes | $N$ | 103 | |
| Amount of Data Collected by One Node in a Data Collection Cycle | $c$ | 1000 | $bit$ |
| Friction | $f$ | 30 | $N$ |
| Adjustable Parameter | $\eta$ | 5400 | |
| Adjustable Parameter | $\lambda$ | 0.23 | |
| Adjustable Parameter | $\beta$ | 2 | |
| Adjustable Parameter | $\alpha$ | 0.5 | |
| Adjustable Parameter | $\varphi$ | 1.7 | |
| Initial Energy of Node | $E_0$ | 2.0 | $J$ |
| Threshold of the Residual Energy | $\delta$ | 0.2 | $J$ |
| Energy Consumption of Sending and Receiving Circuit | $E_{elec}$ | 50 | $nJ \times b^{-1}$ |
| Energy Consumption of Amplifier in Free-Space Model | $\varepsilon_{fs}$ | 10 | $pJ \times (b/m^2)^{-1}$ |
| Energy Consumption of Amplifier in Multi-path Fading Model | $\varepsilon_{amp}$ | 0.0013 | $pJ \times (b/m^4)^{-1}$ |

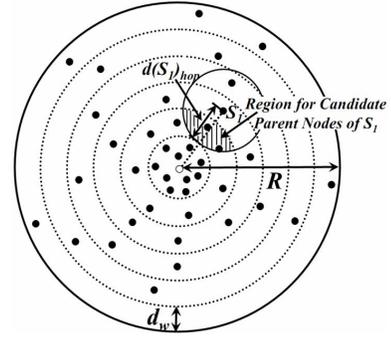

Fig. 13. Node deployment model in SNAA [35].

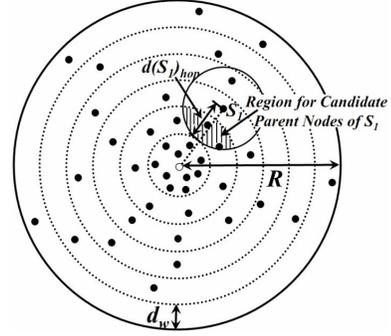

Fig. 14. Data uploading in SNAA [35].

### A. Values of the Coefficients of Virtual Force

It can be seen from Section III that, the value of $\lambda$ and $\eta$ will determine the magnitude of the force to a large extent and ultimately affect the distribution of nodes. Figure 15 shows the results of node distribution with different values of $\lambda$ and $\eta$ after the virtual force is applied. Without loss of generality, it is assumed that 103 nodes are initially randomly deployed in the circular network. To better show the nodes' distribution in different annuluses, boundaries of these annuluses (except the network boundary) are painted with blue lines.

It is not difficult to know from figure 15(a) that, when $\lambda=0.4$ and $\eta=2000$, the nodes' distribution after virtual force adjustment is not ideal. All nodes are clustered near the base station and are located in the two innermost annuluses. According to formula (3), we can see that the virtual force between nodes in this case is mostly gravitation. The magnitude of the repulsive force is close to or equal to the virtual gravity only when the distance between nodes is very short. Therefore, all nodes are close to the center of the network due to the effect of gravitation. What's more, when the distance between each of them is short, the values of virtual gravity and repulsion are equalized with each other so that nodes will ultimately stay at or near the base station.

It can be seen from Figure 15(b) that the distribution of nodes is relatively uniform in the case of $\lambda=0.02$ and $\eta=12000$, but it tends to approach to the network boundary. This is because most of the virtual force between nodes in this case are repulsion. Only when the distance between them is short, gravitation is generated (the repulsion is also large). Therefore, nodes begin to stay away from the network center. Since the boundary repulsive force is adopted in VFEM, nodes can not move out of the network, but there are a considerable number of nodes have moved to the vicinity of the boundary.

From Section III.C-III.E, it is known that, neither of the above cases can ensure the full coverage of the network. That is, the virtual gravitation coefficient $\lambda$ cannot be too small and the virtual repulsion coefficient $\eta$ cannot be too large. By executing a mass of experiments, it is found that, distribution of nodes after virtual force adjustment is ideal when $\eta=5400$ and $\lambda=0.23$, as shown in Figure 15(c).

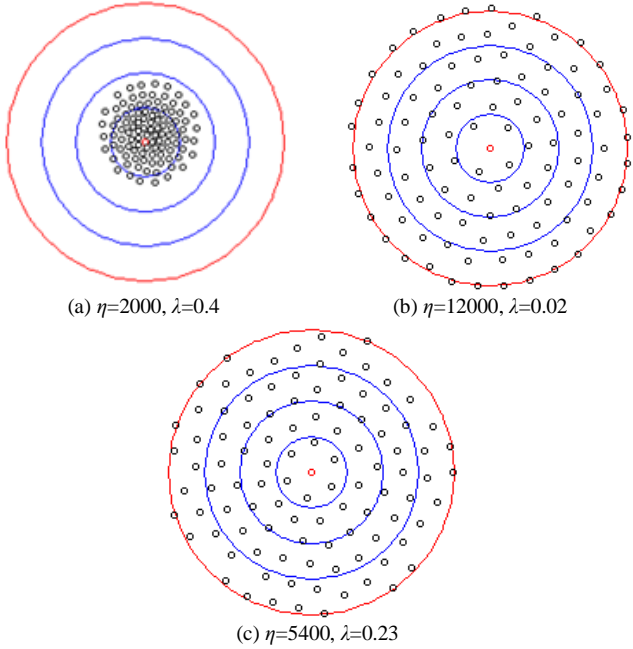

(a) $\eta=2000$, $\lambda=0.4$    (b) $\eta=12000$, $\lambda=0.02$

(c) $\eta=5400$, $\lambda=0.23$
Fig. 15. Distribution of nodes after virtual force adjustment.

### B. Number of Nodes in Each Annulus

The mathematical expectations of the hop distance between two nodes in adjacent annuluses are shown in Table II. Values of these expectations are calculated by formula (15) when $R=100$ and $k$ takes 4, 5 and 6, respectively. It is not difficult to find that when the value of $k$ is unchanged, the hop distance expectations of nodes are substantially equal to each other except for the innermost annulus. So, for most of the relay nodes, energy consumption on uploading one bit of data to their next-hop successor are nearly the same. Although the closer to the center of the network, the more the amount of data to be uploaded by the relay nodes, the number of relay nodes located in the vicinity of the network center is larger than that located in other annuluses (as shown in Table III and Table IV). So, total energy consumption of nodes in each annulus are basically the same, which is consistent with the requirements of formula (7).

TABLE II
MATHEMATICAL EXPECTATIONS OF THE HOP DISTANCE
BETWEEN TWO NODES IN ADJACENT ANNULUSES

|  | $k=4$ | $k=5$ | $k=6$ |
|---|---|---|---|
| $d_0$ | 12.5 m | 10 m | 8.3333 m |
| $d_1$ | 29.8094 m | 23.8475 m | 19.8729 m |
| $d_2$ | 29.5595 m | 23.6476 m | 19.7063 m |
| $d_3$ | 29.5309 m | 23.6247 m | 19.6873 m |
| $d_4$ | - | 23.6171 m | 19.6810 m |
| $d_5$ | - | - | 19.6781 m |

Experimental results of the number of the sensing and relay nodes in each annulus are shown in Table III and IV. It can be found out that, total number of nodes is increasing from the outermost annulus to the innermost one, but the increasing rate has been slowing down. Total number of nodes in $C_1$ is very close to that in $C_0$. So, it can be concluded that VFEM can solve the energy-hole problem by the optimal distribution of nodes based on virtual force generated from annulus. In addition, the number of sensing nodes in each annulus is small, which not only ensures the full coverage of the whole network but also minimizes the amount of the redundant data.

TABLE III
TOTAL NUMBER OF SENSING AND RELAY NODES IN EACH ANNULUS ($N=103$, $k=4$)

| $C_m$ | $N_m^s$ | $N_m^r$ | Total number of Nodes |
|---|---|---|---|
| $C_0$ | 1 | 35 | 36 |
| $C_1$ | 4 | 30 | 34 |
| $C_2$ | 7 | 17 | 24 |
| $C_3$ | 9 | 0 | 9 |

TABLE IV
TOTAL NUMBER OF SENSING AND RELAY NODES IN EACH ANNULUS ($N=200$, $k=5$)

| $C_m$ | $N_m^s$ | $N_m^r$ | Total number of Nodes |
|---|---|---|---|
| $C_0$ | 1 | 57 | 58 |
| $C_1$ | 4 | 52 | 56 |
| $C_2$ | 7 | 38 | 45 |
| $C_3$ | 9 | 21 | 30 |
| $C_4$ | 11 | 0 | 11 |

### C. Residual Energy of Nodes in Each Annulus

Figure 16 and 17 show the residual energy of nodes in each annulus at different network running times. Time of rounds at the end of the network lifetime are 28956 and 34277, respectively. The residual energy of each node in the outermost annulus of the network at the same moment is basically the same in any case. This is because in VFEM, nodes in the outermost annulus are all sensing nodes and they only need to upload their own sensed data to the next-hop neighbor. As mentioned before, nodes in the outermost annulus have the same amount of data to be uploaded in the same period. Furthermore, it can be seen from Section III.E that the lengths of single hop distances between nodes are nearly the same with each other. So, lines in Figure 16(d) and Figure 17(e) are substantially smooth.

Similarly, energy consumption of nodes in the innermost annulus are also basically the same. This is because in VFEM, most of nodes in the innermost annulus are relay nodes and the hop distance for data uploading is only $d_w/2$. Moreover, it is known from Section III.D that, these relay nodes are uniformly distributed in the center of the annulus, so it is almost no difference among them on energy consumption.

Except for the innermost and the outermost annuluses, the difference on energy consumption of nodes in other annuluses is a little large. Residual energy of some nodes is too small. According to the data collection path establishment method in VFEM, it can be seen that not all nodes with forwarding function are selected as the relay in the data upload path when the network starts to run. Thus, the energy of some nodes can be preserved at the beginning of the network lifetime, which causes that the residual energy of nodes are different with each other. However, it is not difficult to see from Figure 16(b), 16(c), 17(b), 17(c) and Figure 17(d) that at the end of the network lifetime, the difference of the residual energy of the nodes in these annuluses is not significant (e.g., the pink lines in these figures). As the difference on energy consumption between nodes continues to increase, some relay nodes are unable to undertake the data uploading task, so the sensing node are likely to reconstruct the routing path according to formula (19). This realizes the energy balance between nodes, and it effectively postpones the generation time of energy-hole.

TABLE V
PERCENTAGE OF RESIDUAL ENERGY FOR NODES IN EACH ANNULUS

|  |  | 5000 rounds | 10000 rounds | 15000 rounds |
|---|---|---|---|---|
| $C_0$ | $N$=103 | 85.53% | 71.07% | 56.61% |
|  | $N$=200 | 86.28% | 72.57% | 58.85% |
| $C_1$ | $N$=103 | 85.84% | 71.68% | 57.51% |
|  | $N$=200 | 86.50% | 73.00% | 59.49% |
| $C_2$ | $N$=103 | 85.90% | 71.77% | 57.64% |
|  | $N$=200 | 86.34% | 72.67% | 59.00% |
| $C_3$ | $N$=103 | 85.70% | 71.19% | 56.54% |
|  | $N$=200 | 86.38% | 72.75% | 59.11% |
| $C_4$ | $N$=103 | - | - | - |
|  | $N$=200 | 86.36% | 72.45% | 58.50% |

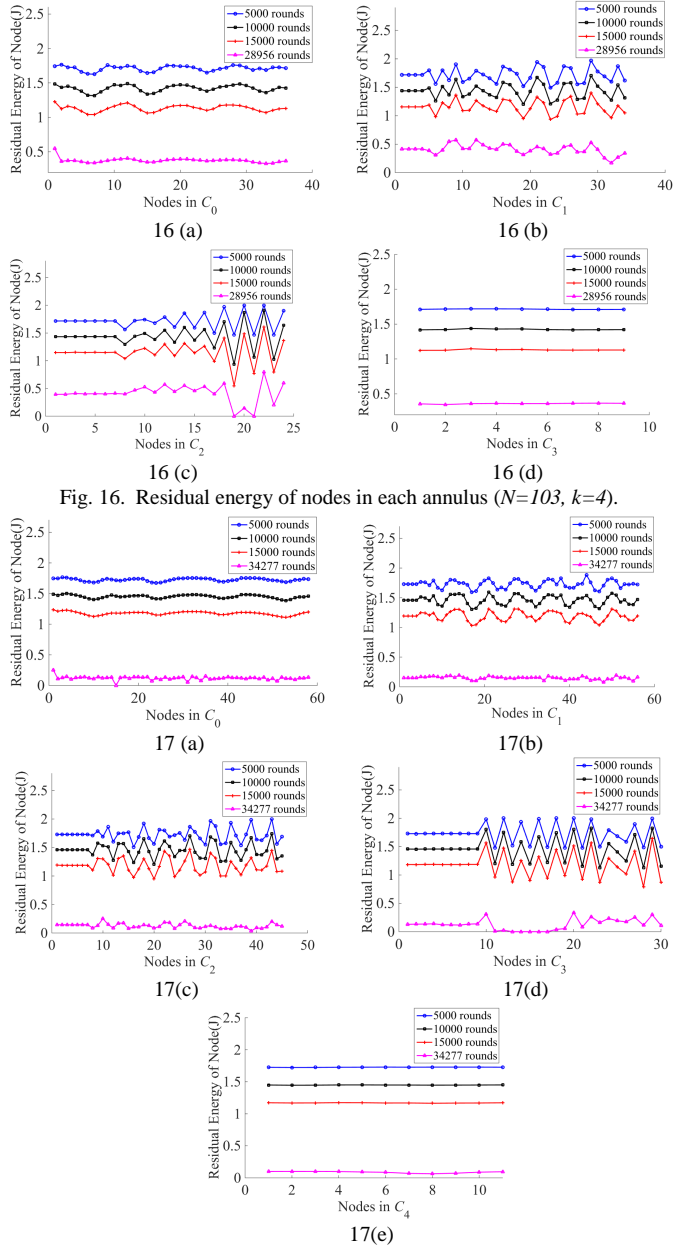

Fig. 16. Residual energy of nodes in each annulus ($N=103$, $k=4$).

Fig. 17. Residual energy of nodes in each annulus ($N=200$, $k=5$).

Table V shows the total residual energy of nodes in each annulus as the percentage of the initial total energy of them. From formula (7), it is known that, this percentage reflects the energy consumption rate of each node within the annulus.

It can be seen from Table V that, regardless of the number of nodes in the network, energy consumption rate of the nodes in each annulus are nearly the same at all time points of the entire network lifetime, which shows the balance of energy consumption. This is because VFEM optimizes the load on each node with the help of the virtual force between sensors as well as the virtual force generated from annulus. It further balances energy consumption between the relay nodes in the same annulus due to the optimal path for data uploading.

### D. Comparison about the Virtual Annulus based Energy-hole Mitigation Strategies

#### 1) Residual Energy of Nodes

Percentages of the total residual energy of nodes in the whole network are shown in Figure 18 and 19. For the networks with different scales, the lifetime of them are also different from each other (28956 rounds and 34277 rounds, respectively). Thus, the abscissas in Figure 18 and Figure 19 are unified as the percentage about the total number of rounds of the entire network lifetime. It can be seen from these two figures that the decline rates of the residual energy of all the three algorithms are nearly stable, regardless of the size of the network. Moreover, the proportion of this value is low at the end of the network lifetime. To achieve balance of energy consumption, the non-uniform deployment of nodes is adopted in all the three algorithms. Thus, there are much more nodes located in the annulus near the center, that effectively relieves the burden on nodes near the center.

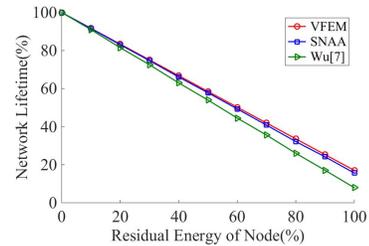

Fig. 18. Total Residual Energy of Nodes ($N=103$, $k=4$).

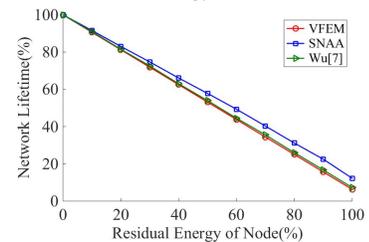

Fig. 19. Total Residual Energy of Nodes ($N=200$, $k=5$).

### 2) Effect on Energy-hole Mitigation under Different Number of Annuluses

It is known from Figure 20 that, with the increase of the number of annuluses, the residual energy at the end of network lifetime is declining in all of the three algorithms. That is, when the network size is larger, the effect on energy mitigation is better. From the second outer annulus, Wu *et al*. uses the method of deploying the inter nodes with equal proportion [7]. If the number of annuluses and nodes is larger, there are more nodes near the network center which ensures the balance on energy consumption. Table VI shows the number of nodes located in different annuluses in Wu [7], SNAA, and VFEM (total number of annuluses is 6). On the other hand, a sleep scheduling strategy is adopted in SNAA. With the increase of the number of annulus, there are more and more sleeping nodes, which effectively reduce the energy consumption. In VFEM, two types of nodes (sensing nodes and relay nodes) are uniformly deployed in each annulus respectively with the help of two kinds of virtual force. So, energy consumption rates of nodes in different annuluses are basically the same. Thus, balance of energy consumption in VFEM is better than that in Wu [7] and SNAA.

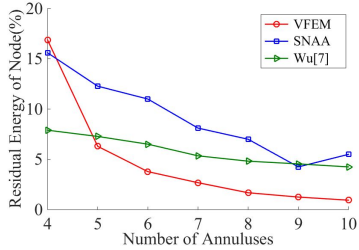

Fig. 20. Effect on Energy-hole Mitigation under Different Number of Annuluses.

Moreover, in SNAA, balance of energy consumption is optimal when the number of annuluses is 9, as shown in Figure 20. However, with the number of annuluses continues to increase, the energy-hole mitigation effect is slightly worse. In SNAA, data is uploaded synchronously, so when the number of annuluses is larger, there are much time spending on data uploading. Moreover, it also increases energy consumption on monitoring. In addition, in the method proposed by Wu [7], the load on nodes near the center of the network can be decreased if the number of nodes increases proportionally from the outer annulus to the inner one. Nevertheless, when the network is divided into more annuluses, the number of nodes in the vicinity of the network center will increase sharply since the ratio setting is not entirely reasonable.

TABLE VI
NUMBER OF NODES IN EACH ANNULUS

|  | VFEM | SNAA | Wu [7] |
|---|---|---|---|
| Number of Nodes in $C_0$ | 84 | 40 | 648 |
| Number of Nodes in $C_1$ | 82 | 33 | 216 |
| Number of Nodes in $C_2$ | 71 | 27 | 72 |
| Number of Nodes in $C_3$ | 56 | 22 | 24 |
| Number of Nodes in $C_4$ | 37 | 18 | 8 |
| Number of Nodes in $C_5$ | 13 | 18 | 4 |

As shown in Table VI, when the number of annuluses is 6, the number of nodes in the innermost annulus is as high as 648. These nodes produce more redundant data, so the balance on energy consumption in Wu [7] is not as good as that in VFEM.

### 3) Energy Consumption Balance about each Node in Different Annulus

The average residual energy about each node in different annuluses is shown in Figure 21. In VFEM, at the end of the network lifetime, the average residual energy of all nodes are basically the same with each other, which shows a good energy balance. However, in SNAA, the average residual energy of nodes in most annuluses is higher than that of VFEM, and the difference of them between annuluses and annuluses is larger. This is because SNAA does not distinguish sensing nodes and relay nodes as VFEM does. If a node $S_i$ is located at the parent node's candidate region of $S_j$, $S_i$ is likely to be a successor to $S_j$. Meanwhile, $S_i$ itself still needs to monitor the surrounding environment. Therefore, effect on energy-hole mitigation in SNAA is not better than in VFEM.

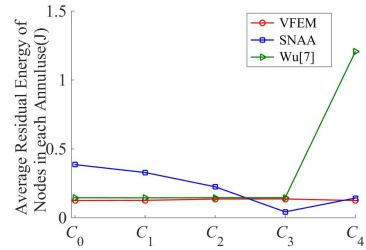

Fig. 21. Comparison of Average Residual Energy for Nodes in Each Annulus.

Furthermore, in Wu's method, except the outermost annulus, energy balance effect about nodes in the rest of the annuluses is similar to that of VFEM. However, at the end of the network lifetime, the average residual energy of nodes in the outermost annulus is higher. In this case, it is impossible to ensure that the energy consumption rate of the nodes in the outermost annulus is consistent with nodes in other annuluses. This is the reason why the last data point value for the green line in Figure 21 is too large.

## V. Conclusion

An energy-hole mitigation method based on two kinds of virtual force is proposed in this paper. With the help of "virtual force between nodes", the network is full covered without blind areas. Subsequently, the work load between nodes is also balanced thanks to the "virtual force generated from annulus". Moreover, we construct the "data forwarding area" for each node, which not only greatly reduces the routing overhead, but also achieves energy balance between different types of nodes to a certain extent. Simulation results show that our method is superior to some typical algorithms in prolonging network lifetime and mitigating energy hole problem.

Although the network topology is optimized in our method, it is undeniable that the number of hops for data transmission will increase in the case of large network scale due to the increase of the number of annuluses. This will inevitably result in high transmission delay. Therefore, improving the real-time performance of the system is one of the goals in our future work.